\documentclass[lettersize,journal,twoside]{IEEEtran}
\usepackage{amsmath,amsfonts}
\usepackage{algorithmic}
\usepackage{algorithm}
\usepackage{array}
\usepackage[caption=false,font=footnotesize,labelfont=rm,textfont=rm]{subfig}
\usepackage{textcomp}
\usepackage{stfloats}
\usepackage{url}
\usepackage{verbatim}
\usepackage{graphicx}
\usepackage{cite}
\usepackage{amsmath}
\usepackage{mathtools}
\usepackage{amssymb}
\usepackage{upgreek}
\usepackage{subfig}
\usepackage{hyperref}
\hypersetup{hypertex=true,
colorlinks=true,
linkcolor=blue,
anchorcolor=blue,
citecolor=blue}
\usepackage{xcolor}
\begin{document}

\title{AN-Aided Secure Beamforming for ELAA-SWIPT \\in Mixed Near- and Far-Field}

\author{
Yaqian Yi, Guangchi Zhang, Miao Cui, Changsheng You,
and Qingqing Wu
\thanks{The work was supported in part by the Guangdong Basic and Applied Basic Research Foundation under Grant 2023A1515011980 and Grant 2024A1515010097; in part by the Shenzhen Science and Technology Program under Grant 20231115131633001 and Grant JCYJ20240813094212016; in part by the Program under Grant 2023QN10X152; in part by NSFC 62371289 and Shanghai Jiao Tong University 2030 Initiative.} 

\thanks{Y. Yi, G. Zhang, and M. Cui are with the School of Information Engineering, Guangdong University of Technology, Guangzhou 510006, China (e-mail: yqyiamanda@163.com; gczhang@gdut.edu.cn; cuimiao@gdut.edu.cn). C. You is with the Department of Electronic and Electrical Engineering, Southern University of Science and Technology, Shenzhen 518055, China (e-mail: youcs@sustech.edu.cn). Q. Wu is with the Department of Electronic Engineering, Shanghai Jiao Tong University, Shanghai 200240, China (e-mail: qingqingwu@sjtu.edu.cn). (Corresponding authors: G. Zhang; M. Cui.) } }

\markboth{IEEE Wireless Communications Letters}
{Yi \MakeLowercase{\textit{et al.}}: AN-Aided Secure Beamforming for ELAA-SWIPT in Mixed Near- and Far-Field}

\IEEEpubid{0000--0000/00\$00.00~\copyright~2025 IEEE}

\maketitle

\begin{abstract}
This letter investigates secure hybrid beamforming (HB) design for an extremely large-scale antenna array-aided simultaneous wireless information and power transfer (SWIPT) system operating in a mixed near-field (NF)/far-field (FF) environment. A base station (BS) employs HB to transmit information and artificial noise (AN) signals simultaneously to multiple FF information receivers (IRs) and NF energy receivers (ERs). The objective is to maximize the weighted sum secrecy rate for the IRs, considering both Type-I (unable to cancel AN) and Type-II (capable of canceling AN) IRs, subject to minimum energy harvesting requirements at the ERs and a BS transmit power constraint. We formulate optimization problems for both IR types and develop an efficient iterative algorithm based on successive convex approximation. Simulation results validate the proposed scheme and provide crucial insights into the security performance of mixed-field SWIPT systems, highlighting the influence of visibility regions and angular user separation.
\end{abstract}


\section{Introduction}
\IEEEPARstart{E}{xtremely} large-scale antenna arrays (ELAAs) are poised to be a cornerstone technology for future sixth-generation networks, offering substantial gains in spectral efficiency and spatial resolution \cite{wang2024a}. A key consequence of the significantly enlarged array aperture in ELAA systems is the expansion of the near-field (NF) region, making mixed NF/far-field (FF) communication scenarios increasingly prevalent and practical \cite{wang2024a}. Extending simultaneous wireless information and power transfer (SWIPT) systems to leverage ELAAs in these mixed-field settings is therefore a natural and promising direction \cite{zhang2024swipt}. This NF energy receiver (ER)/FF information receiver (IR) configuration is particularly relevant: the expanded NF region of ELAAs makes it likely that dedicated ERs, often placed closer for efficient energy transfer, reside in the NF, while primary IRs (e.g., mobile users) may often be located further away in the FF. However, this extension introduces a critical physical layer security (PLS) challenge: information signals intended for legitimate FF IRs can more easily leak to potentially untrusted NF ERs, even when users are angularly separated, due to the complex NF propagation characteristics \cite{zhang2023mixed}. Techniques such as artificial noise (AN) and secure beamforming are thus essential to mitigate this information leakage risk.

While prior work has explored ELAA-SWIPT, focusing on aspects like transmit power minimization using hybrid beamforming (HB) \cite{zhang2024simultaneous} or energy harvesting maximization via resource allocation \cite{zhang2024swipt}, the specific security challenges inherent in the mixed-field context remain unaddressed. Furthermore, although various techniques, including reconfigurable intelligent surfaces \cite{zhao2023secrecy}, AN \cite{zhou2019secure, alageli2018optimization}, and beamforming \cite{zhao2023secrecy, zhou2019secure, alageli2018optimization}, have been employed to enhance PLS performance. However, these studies assume purely FF environments. They consequently overlook the unique security implications arising from mixed NF/FF propagation and the associated channel models (e.g., spherical wavefronts, visibility regions (VRs)).

Motivated by these gaps, this letter investigates AN-aided secure beamforming design for an ELAA-SWIPT system operating in a mixed NF/FF environment. We consider a scenario where a base station (BS) employing an HB architecture simultaneously transmits information signals to multiple FF IRs and energy signals (in the form of AN) to multiple NF ERs. Accounting for practical receiver limitations, we analyze two IR receiver types: Type-I (unable to cancel AN) and Type-II (capable of canceling AN). For each receiver type, we formulate a problem to maximize the weighted sum secrecy rate (WSSR) of IRs by jointly optimizing the information and AN beamformers under constraints on the minimum total energy harvested by the ERs and the maximum transmit power at the BS. The resulting non-convex optimization problems are efficiently tackled using an iterative algorithm based on successive convex approximation (SCA). Simulations validate our scheme and offer insights into how IR receiver type, VR size, and IR angle impact mixed-field secrecy.

\section{System Model}
\begin{figure}[!t]
\centering
\includegraphics[width=2.6in]{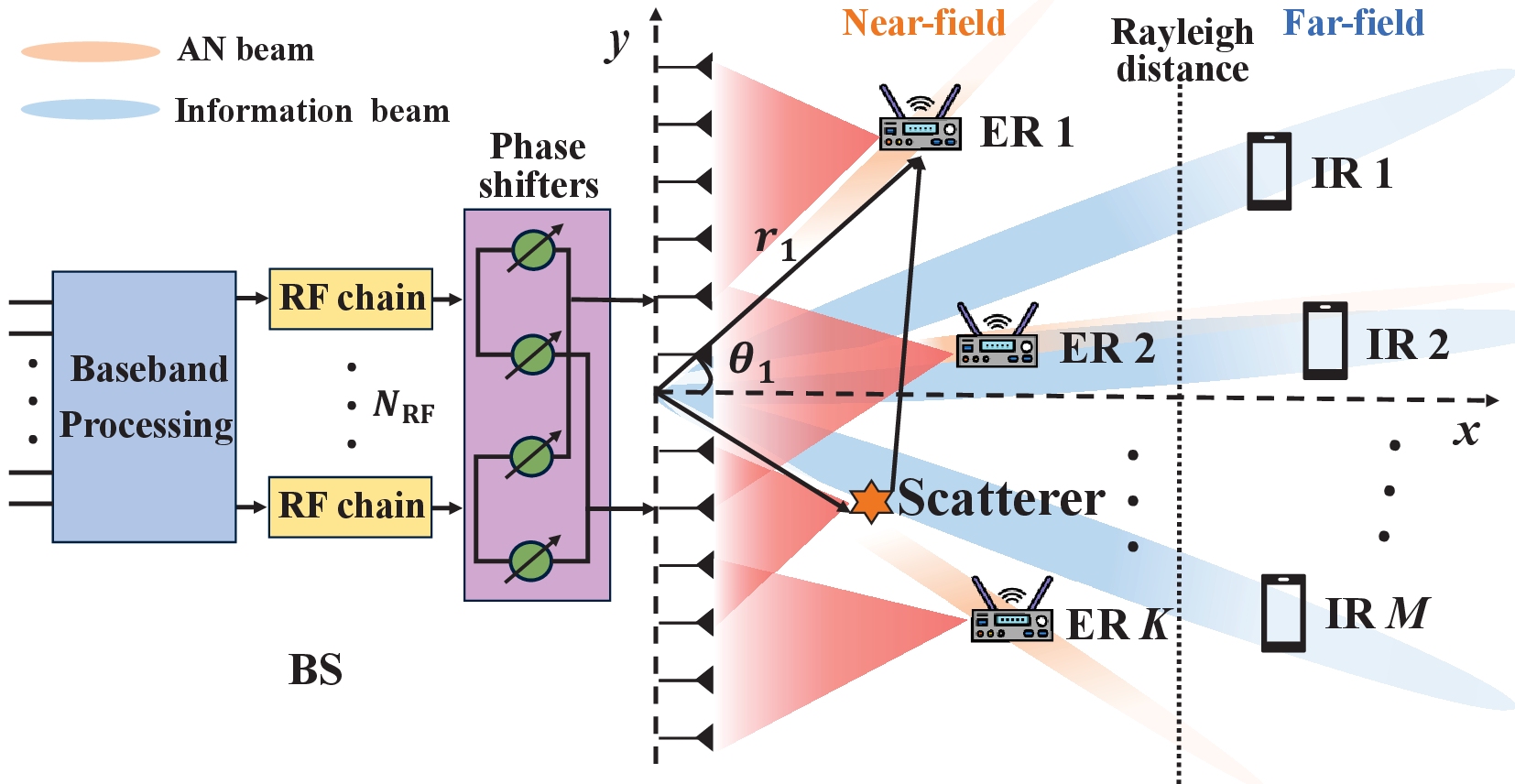}
\caption{An ELAA-SWIPT system with multiple FF IRs and NF ERs.}
\label{fig_1}
\vspace{-0.5cm}
\end{figure}

\IEEEpubidadjcol
We consider an ELAA-SWIPT system in Fig.~\ref{fig_1}, comprising a BS with an $N$-antenna extremely large-scale uniform linear array (ULA),\footnote{Our work can be readily extended to other array architectures, such as the uniform planar array (UPA), by replacing the BS-ERs and BS-IRs channel models with those constructed under the UPA configuration.} $K$ single-antenna ERs, and $M$ single-antenna IRs. A two-dimensional coordinate system is centered at the ULA, with antenna elements along the $y$-axis. The $n$-th element is at $(0, \delta_n d)$, where $\delta_n = \frac{2n-N+1}{2}$ ($n=0, \dots, N-1$) and $d=\lambda_{\text{c}}/2$ is the antenna spacing, with $\lambda_{\text{c}}$ being the carrier wavelength. Since dedicated ERs might be placed closer to the source than primary information users, we consider ERs located in the NF and IRs located in the FF of the BS, respectively \cite{zhang2024swipt}. Specifically, the distances of ERs satisfy $d_{\text{F}} < d < d_{\text{R}}$, while those of IRs satisfy $d > d_{\text{R}}$. Here, $d_{\text{R}}=2D^2/\lambda_{\text{c}}$ (Rayleigh distance) and $d_{\text{F}}=0.62\sqrt{D^3/\lambda_{\text{c}}}$ (Fresnel distance) \cite{wang2024a}, with $D$ being the array aperture. The BS simultaneously transmits information signals to IRs and energy signals to ERs. However, due to wireless broadcast and potential energy spread in mixed NF/FF scenarios, information intended for IRs may leak to ERs,\footnote{We focus on NF eavesdroppers due to their stronger interception capability compared to FF eavesdroppers. The system model can be readily extended to FF eavesdroppers by adapting the channel representation accordingly. Furthermore, investigating ERs as active eavesdroppers is also meaningful and is left for future work.} even if angularly separated \cite{zhang2023mixed}, posing a security risk. 

The BS employs an HB architecture with $N_{\text{RF}} \ll N$ radio frequency (RF) chains. 
To reduce information leakage risk, AN is transmitted \cite{goel2008guaranteeing}. 
Let $\mathbf{s} \in \mathbb{C}^{M\times 1}$ be the information signal vector with the $m$-th entry $s_m \sim \mathcal{CN}(0,1)$, and $\mathbf{q} \in \mathbb{C}^{G\times 1}$ be the AN vector with the $g$-th entry $q_g \sim \mathcal{CN}(0,1)$, where $G$ is the number of AN streams, satisfying $M+G \leq N_{\text{RF}}$. 
The transmitted signal is
\begin{equation}
\mathbf{x} = \mathbf{F}_{\text{A}}\left(\mathbf{Ws}+\mathbf{Vq} \right),
\label{eq:x}
\end{equation}
where $\mathbf{W}=[\mathbf{w}_1,...,\mathbf{w}_M] \in \mathbb{C}^{N_{\text{RF}}\times M}$ and $\mathbf{V}=[\mathbf{v}_1,...,\mathbf{v}_G] \in \mathbb{C}^{N_{\text{RF}}\times G}$ are the information and AN digital beamformers, respectively, and $\mathbf{F}_{\text{A}}\in \mathbb{C}^{N\times N_{\text{RF}}}$ is the analog beamformer.

We consider that the BS has perfect channel state information (CSI) of all IRs and ERs.\footnote{To guarantee the fulfillment of energy harvesting requirement, we assume perfect CSI at the ERs, even when they are treated as potential eavesdroppers. Nevertheless, studying the impact of imperfect CSI at the eavesdroppers on the secrecy performance and investigating the robust algorithm design are meaningful directions and will be considered in our future work.} As such, the received signal at ER $k$ is
\begin{equation}
y_{k}^{\text{EH}}=( \mathbf{h}_{k}^{\text{EH}} ) ^H\mathbf{x}+n_{k}^{\text{EH}},
\label{eq:yeh}
\end{equation} 
where $\mathbf{h}_{k}^{\text{EH}}$ is the BS-ER $k$ channel vector, and $n_{k}^{\text{EH}} \sim \mathcal{CN}(0, \sigma_{\text{EH},k}^2)$ is the additive noise. 
The NF channel $\mathbf{h}_{k}^{\text{EH}}$ is modeled as \cite{chen2024non}
\begin{equation}
\mathbf{h}_{k}^{\text{EH}}=\sqrt{\frac{1}{L_k}}\sum_{l=1}^{L_k}{g_{k,l}\mathbf{b}}\left( \theta _{k,l},r_{k,l} \right) \odot \mathbf{t}\left( \varUpsilon _{k,l} \right) ,
\label{eq:heh}
\end{equation}
where $L_k$ is the number of paths (Line-of-Sight (LoS) for $l=1$, non-LoS (NLoS) for $l>1$), $g_{k,l}$ is the complex gain of the $l$-th path (distance $r_{k,l}$) with $|g_{k,l}| = \frac{\lambda_{\text{c}}}{4\pi r_{k,l}}$, and $\mathbf{b}(\theta_{k,l}, r_{k,l})$ is the NF steering vector for the $l$-th path
\begin{equation}
\boldsymbol{b}\left( \theta _{k,l},r_{k,l} \right) =\left[ e^{-j2\pi ( r_{k,l}^{\left( 0 \right)}-r_{k,l} )/\lambda _{\text{c}}},...,e^{-j2\pi ( r_{k,l}^{\left( N-1 \right)}-r_{k,l} )/\lambda _{\text{c}}} \right] ^T.
\label{eq:b}
\end{equation}
Here, $r_{k,l}^{(n)} = \sqrt{r_{k,l}^2 + (\delta_n d)^2 - 2 r_{k,l} \delta_n d \sin\theta_{k,l}}$ is the distance between the $n$-th antenna element and ER $k$ (if LoS) or scatterer $l$ (if NLoS), and $\theta_{k,l}$ is the path angle.
Due to NF spatial non-stationarity, the antennas of the BS have different visibilities to ERs and scatterers, described by VR, as shown in Fig.~\ref{fig_1}. $\varUpsilon_{k,l} \subseteq \{0, \dots, N-1\}$ is the index set of visible antennas for the $l$-th path to ER $k$. $\mathbf{t}(\varUpsilon_{k,l}) \in \{0,1\}^{N \times 1}$ is the visibility vector, whose $n$-th entry is\footnote{Our work can be extended to practical spatially non-stationary channel scenarios by incorporating realistic VR modeling.}
\begin{equation}
\left[ \mathbf{t}\left( \varUpsilon _{k,l} \right) \right] _n=\begin{cases}
	1,&		n\in \varUpsilon _{k,l}\\
	0,&		n\notin \varUpsilon _{k,l}.\\
\end{cases}
\label{eq:D}
\end{equation} 
The energy harvested at ER $k$ is\footnote{We adopt a linear energy harvesting model by considering the harvested energy is in the linear regime of the ERs. We will extend our work to the scenario of non-linear ERs in future work.} 
\begin{equation}
Q_k=\xi \left( \sum_{m=1}^M{|}( \mathbf{h}_{k}^{\text{EH}} ) ^H\mathbf{F}_{\text{A}}\mathbf{w}_m|^2+\sum_{g=1}^G{|}( \mathbf{h}_{k}^{\text{EH}} ) ^H\mathbf{F}_{\text{A}}\mathbf{v}_g|^2 \right), 
\label{eq:q}
\end{equation}
where $\xi$ is the energy harvesting efficiency. The signal-to-interference-plus-noise ratio (SINR) for potential eavesdropping by ER $k$ on the signal of IR $m$ is
\begin{flalign}
\label{eq:gamma_e} 
&\gamma^\text{e}_{m,k} = \frac{|(\mathbf{h}_{k}^{\text{EH}})^H \mathbf{F}_{\text{A}} \mathbf{w}_m|^2}{\sum\limits_{g=1}\limits^G |(\mathbf{h}_{k}^{\text{EH}})^H \mathbf{F}_{\text{A}} \mathbf{v}_g|^2 + \sum\limits_{j=1, j \neq m}\limits^M |(\mathbf{h}_{k}^{\text{EH}})^H \mathbf{F}_{\text{A}} \mathbf{w}_j|^2 + \sigma_{{\text{EH}},k}^2}. 
\end{flalign}

The received signal at IR $m$ is
\begin{equation}
y_{m}^{\text{ID}}=( \mathbf{h}_{m}^{\text{ID}} ) ^H\mathbf{x}+n_{m}^{\text{ID}},
\label{eq:yid}
\end{equation}
where $\mathbf{h}_{m}^{\text{ID}}$ is the BS-IR $m$ channel vector, and $n_{m}^{\text{ID}} \sim \mathcal{CN}(0, \sigma_{\text{ID},m}^2)$ is the additive noise. 
The FF channel $\mathbf{h}_{m}^{\text{ID}}$ is modeled as
\begin{equation}
\mathbf{h}_{m}^{\text{ID}}=\sqrt{\frac{1}{L_m}}\sum_{l=1}^{L_m}{g_{m,l}\mathbf{a}\left( \theta _{m,l} \right)},
\label{eq:hid}
\end{equation}
where $L_m$ is the number of paths, $g_{m,l}$ is the complex gain of the $l$-th path, and $\mathbf{a}(\theta_{m,l})$ is the FF steering vector for the $l$-th path with angle $\theta_{m,l}$: 
\begin{equation}
\mathbf{a}\left( \theta _{m,l} \right) =\left[ 1,e^{j\pi \sin \theta _{m,l}},...,e^{j\pi \left( N-1 \right) \sin \theta _{m,l}} \right] ^T.
\label{eq:a}
\end{equation}
Following \cite{xu2014multiuser}, we consider two IR receiver types: Type-I (cannot cancel AN) and Type-II (can cancel AN). Their respective SINRs at IR $m$ are
\begin{align}
\gamma_{m}^{\left( \text{I} \right)} &= \frac{\left| ( \mathbf{h}_{m}^{\text{ID}} ) ^H\mathbf{F}_{\text{A}}\mathbf{w}_m \right|^2}{\sum\limits_{g=1}\limits^G{\left| ( \mathbf{h}_{m}^{\text{ID}} ) ^H\mathbf{F}_{\text{A}}\mathbf{v}_g \right|^2}+\sum\limits_{j=1,j\ne m}\limits^M{\left| ( \mathbf{h}_{m}^{\text{ID}} ) ^H\mathbf{F}_{\text{A}}\mathbf{w}_j \right|^2}+\sigma _{\text{ID,}m}^{2}},
\label{eq:sinr1}
\\
\gamma_{m}^{\left( \text{II} \right)} &= \frac{\left| ( \mathbf{h}_{m}^{\text{ID}} ) ^H\mathbf{F}_{\text{A}}\mathbf{w}_m \right|^2}{\sum\limits_{j=1,j\ne m}\limits^M{\left| ( \mathbf{h}_{m}^{\text{ID}} ) ^H\mathbf{F}_{\text{A}}\mathbf{w}_j \right|^2}+\sigma _{\text{ID,}m}^{2}}. \label{eq:sinr2}
\end{align}

The achievable secrecy rates for IR $m$ in bits per second per Hertz (bps/Hz) are\footnote{If the ERs are considered trusted, the secrecy rate expressions reduce to their corresponding information rates, as there is no information leakage.}
\begin{flalign}
&R_m^{(\text{I})} = \left[ \log_2 \left(1 + \gamma_m^{(\text{I})}\right) - \max_{1 \leq k \leq K} \log_2 \left(1 + \gamma^\text{e}_{m,k}\right) \right]^+, \label{eq:r1} \\ 
&R_m^{(\text{II})} = \left[ \log_2 \left(1 + \gamma_m^{(\text{II})}\right) -\max_{1 \leq k \leq K} \log_2 \left(1 + \gamma^\text{e}_{m,k}\right) \right]^+, \label{eq:r2}
\end{flalign}
where $[ \cdot ]^+ \triangleq \max\{\cdot, 0\}$.

This letter focuses on the beamforming design to maximize the WSSR for the IRs, while satisfying the ERs' energy harvesting requirement and the BS's transmit power budget. 
We jointly optimize the analog beamformer $\mathbf{F}_{\text{A}}$ and the digital beamformers $\mathbf{W}$ (information) and $\mathbf{V}$ (AN). 
The WSSR maximization problems for IRs equipped with Type-I and Type-II receivers are respectively formulated as
\begin{subequations}  \label{P1}
\begin{align}
 \max_{\mathbf{F}_{\text{A}}, \mathbf{W}, \mathbf{V}} ~ & \sum_{m=1}^M \alpha_m R_m^{(\text{I})}  \\
\label{sumq} \text{s.t.} \quad~ & \sum_{k=1}^K Q_k \geq Q_0, \\
\label{sump} &\lVert \mathbf{F}_{\text{A}}\mathbf{W} \rVert ^2_F+\lVert \mathbf{F}_{\text{A}}\mathbf{V} \rVert ^2_F\leq P_{\max}, \\
\label{fa1} &| [ \mathbf{F}_{\text{A}} ] _{n,i}|=1,~\forall n,~i, 
\end{align}
\vspace{-0.5cm}
\end{subequations}
\begin{align}  \label{P2}
\max_{\mathbf{F}_{\text{A}}, \mathbf{W}, \mathbf{V}} ~ & \sum_{m=1}^M \alpha_m R_m^{(\text{II})} \quad \text{s.t.} ~ \text{\eqref{sumq}, \eqref{sump}, \eqref{fa1}}. 
\end{align}
Here, $\alpha_m \ge 0$ is the weight for IR $m$'s secrecy rate. Constraint \eqref{sumq} guarantees a minimum total harvested energy $Q_0$ across all ERs. Constraint \eqref{sump} limits the total BS transmit power to $P_{\max}$. Constraint \eqref{fa1} is the unit-modulus constraint imposed on the elements of the analog beamformer $\mathbf{F}_{\text{A}}$.

\section{Proposed Algorithm for Problems \eqref{P1} and \eqref{P2}}
The objective functions and constraints \eqref{sumq}-\eqref{fa1} make problems \eqref{P1} and \eqref{P2} non-convex. Furthermore, the intricate WSSR expressions and the inherent coupling between the digital beamformers ($\mathbf{W}, \mathbf{V}$) and the analog beamformer $\mathbf{F}_{\text{A}}$ make these problems challenging to solve directly.

We first address problem \eqref{P1}. To manage the complexity associated with the HB architecture and decouple the optimization variables, we adopt a strategy where a fixed $\mathbf{F}_{\text{A}}$ is designed first based on channel characteristics, after which $\mathbf{W}$ and $\mathbf{V}$ are optimized. Specifically, inspired by beam steering principles aimed at aligning beams towards users \cite{zhang2024swipt}, the $i$-th column of $\mathbf{F}_{\text{A}}$, denoted by $\left[ \mathbf{F}_{\text{A}} \right] _i$, is constructed as follows:
\begin{flalign}
\left[ \mathbf{F}_{\text{A}} \right] _i=\begin{cases}
	\mathbf{a}\left( \theta _i \right) ,&		1\leq i\leq M\\
	\mathbf{b}\left( \theta _{i-M},r_{i-M} \right) ,&		M+1\leq i\leq M+G\\
	\frac{1}{\left| \mathbf{c} \right|}\odot \mathbf{c} \,\,,&		M+G+1\leq i\leq N_{\text{RF}},
\end{cases}
\label{FA}
\end{flalign}
where $\theta_i$ and $(\theta_{i-M}, r_{i-M})$ represent the angle (and distance for NF) typically associated with the LoS path of the $i$-th IR or $(i-M)$-th ER, respectively. This design strengthens AN and information signal reception at the ERs and IRs, respectively, while the NF beam focusing effect helps suppress AN interference to FF IRs. The symbol $\odot$ denotes the Hadamard product, and $\mathbf{c}=\sum_{m=1}^M{\mathbf{a}\left( \theta _m \right)}$. The third case in \eqref{FA} utilizes an element-wise normalized version of $\mathbf{c}$ to potentially enhance array gain across multiple IRs \cite{zhang2024simultaneous} while adhering to the unit-modulus requirement per RF chain column.
With $\mathbf{F}_{\text{A}}$ determined by \eqref{FA}, the objective function and constraint \eqref{sumq} remain non-convex functions of $\mathbf{W}$ and $\mathbf{V}$, making the optimization still challenging. According to \cite{cui2019secure}, the non-smooth $[ \cdot ]^+$ operator in the objective function \eqref{eq:r1} can be removed without affecting the optimal solution, assuming the optimal secrecy rate is non-negative. 

To optimize $\mathbf{W}$ and $\mathbf{V}$, we define the effective channel matrices $\mathbf{\tilde{H}}_{m}^{\text{ID}} \triangleq \mathbf{F}_{\text{A}}^{H}\mathbf{h}_{m}^{\text{ID}}( \mathbf{h}_{m}^{\text{ID}} ) ^H\mathbf{F}_{\text{A}}$ and $\mathbf{\tilde{H}}_{k}^{\text{EH}}\triangleq\mathbf{F}_{\text{A}}^{H}\mathbf{h}_{k}^{\text{EH}}( \mathbf{h}_{k}^{\text{EH}} ) ^H\mathbf{F}_{\text{A}}$, as well as the following quadratic terms 
$A_{m} \triangleq \sum_{g=1}^G{\mathbf{v}_{g}^{H}\mathbf{\tilde{H}}_{m}^{\text{ID}}\mathbf{v}_g}$,
$B_{m} \triangleq \sum_{m'=1}^M{\mathbf{w}_{m'}^{H}\mathbf{\tilde{H}}_{m}^{\text{ID}}\mathbf{w}_{m'}}$,
$C_{m} \triangleq \sum_{j=1,j\ne m}^M{\mathbf{w}_{j}^{H}\mathbf{\tilde{H}}_{m}^{\text{ID}}\mathbf{w}_j}$, 
$E_{k} \triangleq \sum_{g=1}^G{\mathbf{v}_{g}^{H}\mathbf{\tilde{H}}_{k}^{\text{EH}}\mathbf{v}_g}$, 
$F_{k} \triangleq \sum_{m=1}^M{\mathbf{w}_{m}^{H}\mathbf{\tilde{H}}_{k}^{\text{EH}}\mathbf{w}_m}$, and 
$G_{m,k} \triangleq \sum_{j=1,j\ne m}^M{\mathbf{w}_{j}^{H}\mathbf{\tilde{H}}_{k}^{\text{EH}}\mathbf{w}_j}$.
We introduce the slack variables $\boldsymbol{\lambda }=[\lambda_1,...,\lambda_M]^T$, $\boldsymbol{\mu }=[\mu_1,...,\mu_M]^T$, $\boldsymbol{\tau }=[\tau_1,...,\tau_K]^T$, and $\boldsymbol{\kappa}=[\boldsymbol{\kappa}_1,...,\boldsymbol{\kappa}_K] $ where the $k$-th column is $\boldsymbol{\kappa}_k=[\kappa_{1,k},...,\kappa_{M,k}]^T$. 
With the definitions and slack variables, problem \eqref{P1} can be reformulated as
\begin{subequations}  \label{P1.1}
\begin{flalign}
\max_{\substack{\mathbf{W}, \mathbf{V}, \boldsymbol{{\lambda}}, \boldsymbol{{\mu}}, \\ \boldsymbol{\tau }, \boldsymbol{\kappa}}}  & \sum_{m=1}^M{\alpha _m}\left( \lambda _m-\mu _m-\max_{1\leq k\leq K}\left( \tau _k-\kappa _{m,k} \right) \right)  \\
\text{s.t.} \quad~ &A_{m}+B_{m}+\sigma _{\text{ID,}m}^{2}\ge 2^{\lambda _m},\quad \forall m,
\label{p1re1}\\
&A_{m}+C_{m}+\sigma _{\text{ID,}m}^{2}\le 2^{\mu _m},\quad \forall m,
\label{p1re2}\\
&E_{k}+F_{k}+\sigma _{\text{EH,}k}^{2}\le 2^{\tau _k},\quad\forall k, \label{p1re3}\\
&E_{k}+G_{m,k}+\sigma _{\text{EH,}k}^{2}\ge 2^{\kappa _{m,k}},\quad \forall m,~k,
\label{p1re4}\\
&\xi \textstyle\sum_{k=1}^K (E_k + F_k) \ge Q_0, \label{sumq_re}\\ 
&\lVert \mathbf{F}_{\text{A}}\mathbf{W} \rVert ^2_F+\lVert \mathbf{F}_{\text{A}}\mathbf{V} \rVert ^2_F\leq P_{\max}. \label{sump_re} 
\end{flalign}
\end{subequations}
Problems \eqref{P1} (with fixed $\mathbf{F}_{\text{A}}$, removed $[\cdot]^+$) and \eqref{P1.1} are equivalent. This equivalence holds because, at the optimal solution of \eqref{P1.1}, constraints \eqref{p1re1}-\eqref{p1re4} must be active (i.e., hold with equality). If \eqref{p1re1} or \eqref{p1re4} were slack for some index, the corresponding $\lambda_m$ or $\kappa_{m,k}$ could be increased; if \eqref{p1re2} or \eqref{p1re3} were slack, the corresponding $\mu_m$ or $\tau_k$ could be decreased. Either action would strictly improve the objective function value without violating other constraints, contradicting the assumption of optimality. Therefore, the slack variables effectively represent the logarithmic terms in the original secrecy rate expression at the optimum.

Problem \eqref{P1.1} remains non-convex due to the constraints \eqref{p1re1}-\eqref{sumq_re}. We propose an iterative algorithm based on the SCA technique to find a locally optimal solution. In each iteration (indexed by $t$), we replace the non-convex constraints with convex approximations derived using the solution obtained from the previous iteration.
Let $\mathbf{w}_{m}^{\left( t \right)}$, $\mathbf{v}_{g}^{\left( t \right)}$, $\mu _{m}^{\left( t \right)}$, and $\tau _{k}^{\left( t \right)}$ denote the feasible points obtained at the $t$-th iteration.

First, consider constraints \eqref{p1re1}, \eqref{p1re4}, and \eqref{sumq_re}, which involve convex quadratic terms ($A_m, B_m, E_k, F_k, G_{m,k}$) on the left-hand side (LHS) of the inequalities. We apply the first-order Taylor expansion to obtain affine lower bounds for these terms \cite{co}. Specifically, we have $A_{m} \ge A_{m}^{\text{lb}(t)}$, $B_{m} \ge B_{m}^{\text{lb}(t)}$, $E_{k} \ge E_{k}^{\text{lb}(t)}$, $F_{k} \ge F_{k}^{\text{lb}(t)}$, and $G_{m,k} \ge G_{m,k}^{\text{lb}(t)}$, where
\begin{align*}
A_{m}^{\text{lb}(t)} &\triangleq \sum_{g=1}^G \left( 2\text{Re}\big\{ ( \mathbf{v}_{g}^{( t)} ) ^H\mathbf{\tilde{H}}_{m}^{\text{ID}}\mathbf{v}_g \big\} - ( \mathbf{v}_{g}^{\left( t \right)} ) ^H\mathbf{\tilde{H}}_{m}^{\text{ID}}\mathbf{v}_{g}^{\left( t \right)} \right), \\
B_{m}^{\text{lb}(t)} &\triangleq \sum_{m'=1}^M \left( 2\text{Re}\big\{ ( \mathbf{w}_{m'}^{\left( t \right)} ) ^H\mathbf{\tilde{H}}_{m}^{\text{ID}}\mathbf{w}_{m'} \big\} -( \mathbf{w}_{m'}^{\left( t \right)} ) ^H\mathbf{\tilde{H}}_{m}^{\text{ID}}\mathbf{w}_{m'}^{(t)} \right), \\
E_{k}^{\text{lb}(t)} &\triangleq \sum_{g=1}^G \left( 2\text{Re}\big\{ ( \mathbf{v}_{g}^{\left( t \right)}) ^H\mathbf{\tilde{H}}_{k}^{\text{EH}}\mathbf{v}_g \big\} - ( \mathbf{v}_{g}^{\left( t \right)}) ^H\mathbf{\tilde{H}}_{k}^{\text{EH}}\mathbf{v}_{g}^{\left( t \right)} \right), \\
F_{k}^{\text{lb}(t)} &\triangleq \sum_{m=1}^M \left( 2\text{Re}\big\{ ( \mathbf{w}_{m}^{\left( t \right)} ) ^H\mathbf{\tilde{H}}_{k}^{\text{EH}}\mathbf{w}_m \big\} - ( \mathbf{w}_{m}^{\left( t \right)} ) ^H\mathbf{\tilde{H}}_{k}^{\text{EH}}\mathbf{w}_{m}^{\left( t \right)} \right), \\
G_{m,k}^{\text{lb}(t)} &\triangleq \sum_{j=1,j\ne m}^M \left( 2\text{Re}\big\{ ( \mathbf{w}_{j}^{\left( t \right)} ) ^H\mathbf{\tilde{H}}_{k}^{\text{EH}}\mathbf{w}_j \big\} -( \mathbf{w}_{j}^{\left( t \right)} ) ^H\mathbf{\tilde{H}}_{k}^{\text{EH}}\mathbf{w}_{j}^{(t)} \right).
\end{align*}
Substituting these lower bounds into \eqref{p1re1}, \eqref{p1re4}, and \eqref{sumq_re} yields the following convex constraints, which tighten the feasible region compared to the original constraints:
\begin{align}
&A_{m}^{\text{lb}(t)}+ B_{m}^{\text{lb}(t)}+\sigma _{\text{ID,}m}^{2}\ge 2^{\lambda _m},\quad \forall m, \label{p1re1_sca}\\
&E_{k}^{\text{lb}(t)}+G_{m,k}^{\text{lb}(t)}+\sigma _{\text{EH,}k}^{2}\geq 2^{\kappa _{m,k}},\quad \forall m,~k, \label{p1re4_sca}\\
&\xi \textstyle\sum_{k=1}^K \left( E_{k}^{\text{lb}(t)} + F_{k}^{\text{lb}(t)} \right) \geq Q_0. \label{q_sca}
\end{align}

Next, consider constraints \eqref{p1re2} and \eqref{p1re3}, which have the form `convex function $\le$ convex function'. We handle the convex exponential functions $2^{\mu_m}$ and $2^{\tau_k}$ on the right-hand side (RHS) by applying the first-order Taylor expansion at the points $\mu_{m}^{(t)}$ and $\tau_{k}^{(t)}$, respectively. This yields affine lower bounds for the RHS functions:
\begin{align*}
\psi_m ^{\text{lb}(t)} &\triangleq 2^{\mu _{m}^{\left( t \right)}}+2^{\mu _{m}^{\left( t \right)}}\ln\text{2\,}\big( \mu _m-\mu _{m}^{\left( t \right)} \big), \\ 
\varphi_k ^{\text{lb}(t)} &\triangleq 2^{\tau _{k}^{\left( t \right)}}+2^{\tau _{k}^{\left( t \right)}}\ln\text{2\,}\big( \tau _k- \tau _{k}^{\left( t \right)} \big). 
\end{align*}
Replacing the RHS of \eqref{p1re2} and \eqref{p1re3} with these affine lower bounds results in the following convex constraints:
\begin{align}
A_{m}+C_{m}+\sigma _{\text{ID,}m}^{2} & \le \psi_m ^{\text{lb}(t)}, \quad \forall m, \label{p1re2_sca}\\
E_{k}+F_{k}+\sigma _{\text{EH,}k}^{2} & \le \varphi_k ^{\text{lb}(t)}, \quad \forall k. \label{p1re3_sca}
\end{align}

By substituting the original non-convex constraints \eqref{p1re1}-\eqref{sumq_re} with their convex approximations \eqref{p1re1_sca}-\eqref{p1re3_sca}, we obtain the following problem to be solved at the $(t+1)$-th iteration:
\begin{subequations}   \label{P1.2}
\begin{flalign} 
\max_{\substack{\mathbf{W}, \mathbf{V}, \boldsymbol{{\lambda}}, \boldsymbol{{\mu}}, \\ \boldsymbol{\tau }, \boldsymbol{\kappa}}}  & \sum_{m=1}^M{\alpha _m}\left( \lambda _m-\mu _m-\max_{1\leq k\leq K}\left( \tau _k-\kappa _{m,k} \right) \right) \\
\text{s.t.} \quad~ &\eqref{sump_re},~\eqref{p1re1_sca},~\eqref{p1re4_sca},~\eqref{q_sca},~\eqref{p1re2_sca},~\eqref{p1re3_sca}.
\end{flalign}
\end{subequations}
Problem \eqref{P1.2} is a convex optimization problem and can be efficiently solved using the interior-point method. The SCA framework ensures that the sequence of objective values obtained by iteratively solving \eqref{P1.2} is non-decreasing and converges to a Karush-Kuhn-Tucker point of the reformulated problem \eqref{P1.1}, which is generally a locally optimal solution \cite{co}. In the algorithm for solving problem \eqref{P1.2}, the variables $\mu _m$ and $\tau _k$ are respectively initialized as $\mu _m^{(0)}=\log _2(A_{m}^{(0)}+C_{m}^{(0)}+\sigma _{\text{ID,}m}^{2} )$ and $\tau _k^{(0)}=\log _2(E_{k}^{(0)}+F_{k}^{(0)}+\sigma _{\text{EH,}k}^{2} )$, where  $A_{m}^{(0)}$, $C_{m}^{(0)}$, $E_{k}^{(0)}$ and $F_{k}^{(0)}$ are the initial values determined by the initial digital beamformers.

Next, we address problem \eqref{P2}. The solution procedure mirrors that for problem \eqref{P1}. We fix $\mathbf{F}_{\text{A}}$ using \eqref{FA} and remove the $[\cdot]^+$ operator from \eqref{eq:r2}. We introduce new slack variables $\boldsymbol{\tilde{\lambda}}=[\tilde{\lambda} _1,...,\tilde{\lambda} _M]^T$ and $\boldsymbol{\tilde{\mu} }=[\tilde{\mu }_1,...,\tilde{\mu }_M]^T$, with $\tilde{\lambda} _m=\log _2\left(B_{m}+\sigma _{\text{ID,}m}^{2} \right)$ and $\tilde{\mu }_m=\log _2\left(C_{m}+\sigma _{\text{ID,}m}^{2} \right)$.
We then apply the SCA-based iterative method. Let $\tilde{\mu }_m^{\left( t \right)}$ and $\tau_k^{(t)}$ be the feasible points from the $t$-th iteration. For the convex LHS in the constraints involving $\tilde{\lambda}_m$ and $\kappa_{m,k}$, and the convex RHS $2^{\tau_k}$ in the constraint related to $\tau_k$, we employ their affine lower bounds. For the constraint related to $\tilde{\mu}_m$, we use the affine lower bound $\phi_m^{\text{lb}(t) } \triangleq 2^{\tilde{\mu }_m^{\left( t \right)}}+2^{\tilde{\mu }_m^{\left( t \right)}}\ln\text{2\,}\big( \tilde{\mu }_m-\tilde{\mu }_m^{\left( t \right)} \big)$ for the RHS $2^{\tilde{\mu}_m}$. 
The resulting problem to be solved at the $(t+1)$-th iteration for problem \eqref{P2} is formulated as
\begin{subequations}  \label{P2.1}
\begin{flalign}
\max_{\substack{\mathbf{W}, \mathbf{V}, \boldsymbol{\tilde{\lambda}}, \boldsymbol{\tilde{\mu}}, \\ \boldsymbol{\tau }, \boldsymbol{\kappa}}}  & \sum_{m=1}^M{\alpha _m}\left( \tilde{\lambda} _m-\tilde{\mu} _m-\max_{1\leq k\leq K}\left( \tau _k-\kappa _{m,k} \right) \right)  \\
\text{s.t.} \quad~ 
&B_{m}^{\text{lb}(t)}+\sigma _{\text{ID,}m}^{2}\geq 2^{\tilde{\lambda} _m},\quad \forall m, \label{p2re1_sca}\\
&C_{m}+\sigma _{\text{ID,}m}^{2} \le \phi_m^{\text{lb}(t) } ,\quad \forall m, \label{p2re2_sca}\\
&\eqref{sump_re}, ~\eqref{p1re4_sca}, ~\eqref{q_sca}, ~\eqref{p1re3_sca}.  
\end{flalign}
\end{subequations}
Similarly, problem \eqref{P2.1} is convex and can be solved efficiently in each iteration. In the algorithm for solving problem (25), the variable $\tilde{\mu}_m$ is initialized according to its definition.

Finally, we discuss the convergence and computational complexity of the proposed iterative algorithms. For both \eqref{P1.1} and \eqref{P2.1}, the SCA-based algorithms generate a sequence of feasible solutions. Since the objective function value is non-decreasing in each iteration and is bounded above due to the transmit power constraint \eqref{sump_re}, the algorithms are guaranteed to converge to a stationary point of the respective reformulated problems \cite{co}. The computational complexity per iteration is primarily determined by solving the convex problems \eqref{P1.2} or \eqref{P2.1}. Therefore, the complexity is approximately $\mathcal{O}\big( I\left( N_{\text{RF}}(M+G)+MK \right) ^{3.5} \big)$, where $I$ denotes the number of iterations required for convergence \cite{rafieifar2023secure}. 

\section{Simulation Results}
\begin{figure*}[!t] 
	\setlength{\abovecaptionskip}{-3pt} 
	\setlength{\belowcaptionskip}{-10pt}
	\centering
	\begin{minipage}[t]{0.33\linewidth}
		\centering
		\includegraphics[width=2in]{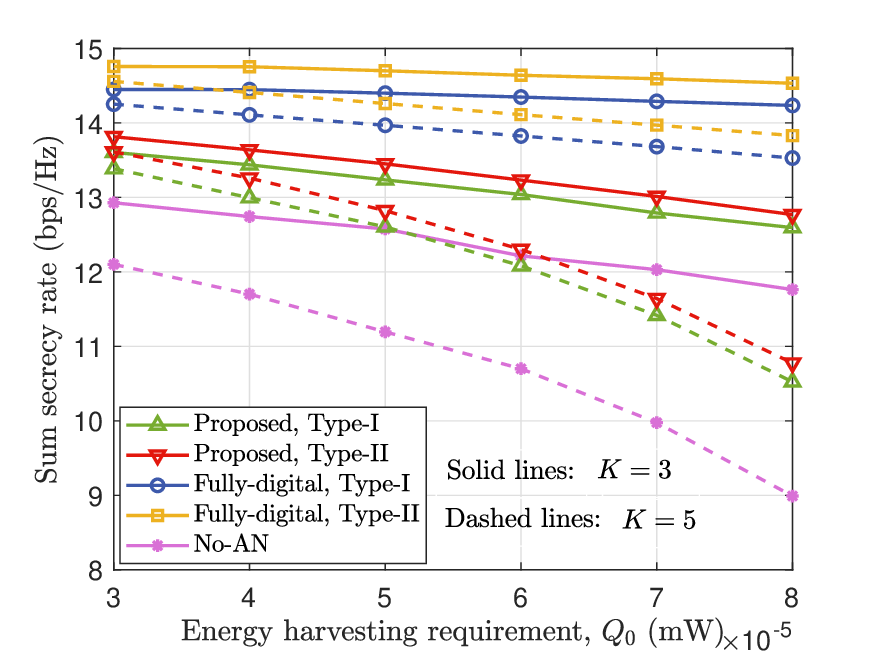}
		\caption{SSR vs. $Q_0$ for different schemes ($K=\protect\\3, 5$).}
		\label{fig2}
	\end{minipage}
	\begin{minipage}[t]{0.33\linewidth}
		\centering
		\includegraphics[width=2in]{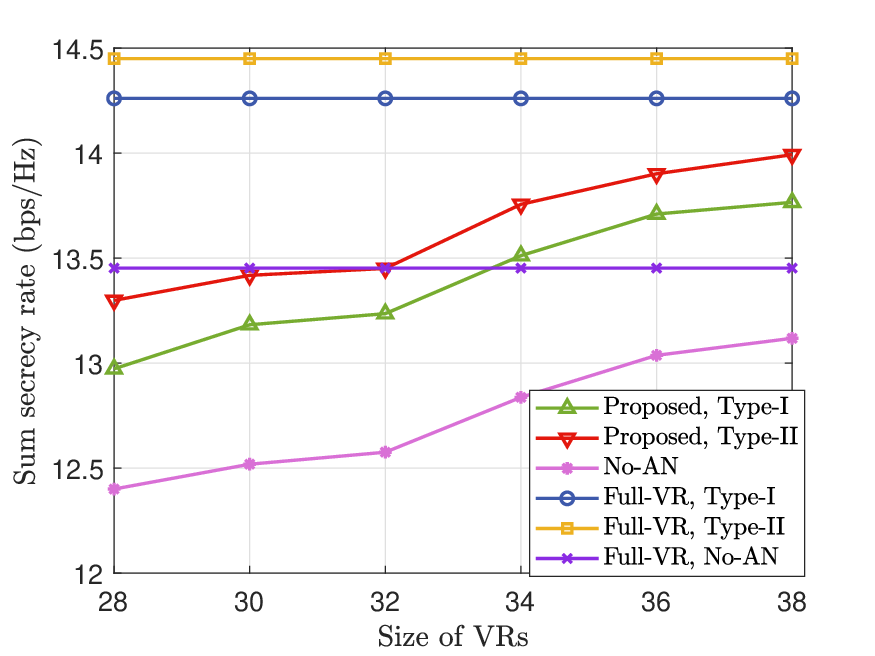}
		\caption{SSR vs. VR size for different schemes.}
		\label{fig3}
	\end{minipage}
	\begin{minipage}[t]{0.33\linewidth}
		\centering
		\includegraphics[width=2in]{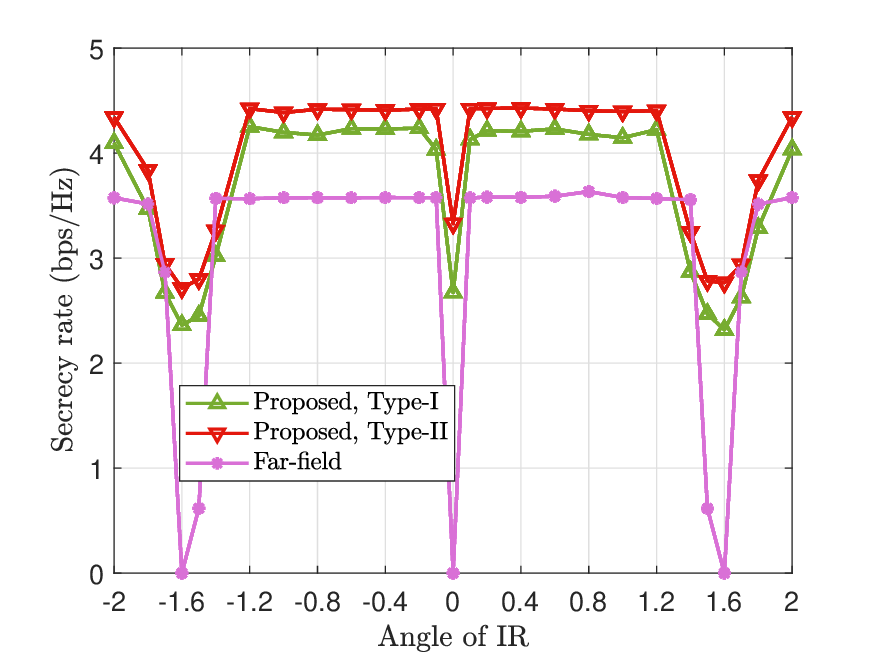}
		\caption{Secrecy rate vs. IR angle.}
		\label{fig4}
	\end{minipage}
\vspace{-0.5cm}
\end{figure*}
This section presents simulation results validating the effectiveness of the proposed scheme and providing insights into the impact of IR receiver types, VR size, and user geometry on secrecy performance in mixed NF/FF SWIPT systems.
Unless otherwise specified, the BS has $N = 128$ antennas and $N_{\text{RF}} = 10$ RF chains, serving $K = 3$ ERs and $M = 2$ IRs. ER locations $(\theta_k, r_k)$ are $(1.3 \text{ rad}, 0.25d_{\text{R}})$, $(0 \text{ rad}, 0.1d_{\text{R}})$, $(-1.1 \text{ rad}, 0.3d_{\text{R}})$. IR locations $(\theta_m, r_m)$ are $(1 \text{ rad}, 1.1d_{\text{R}})$, $(-0.2 \text{ rad}, 1.3d_{\text{R}})$. NF/FF scatterers are randomly distributed in angle at fixed distances $0.1d_{\text{R}}$ and $1.1d_{\text{R}}$, respectively. Key parameters include: $f_{\text{c}}=30\ \text{GHz}$, $P_{\max}=1\ \text{W}$, $\xi =0.5$, $Q_0 = 0.05\ \upmu \text{W}$, noise power $\sigma _{\text{ID,}m}^{2}=\sigma _{\text{EH,}k}^{2}=-80\ \text{dBm}$, $G=K$, $L_k=2$, $L_m=3$, and $\alpha_m = 1$, $\forall m, k$. The digital beamformers $\mathbf{W}$ and $\mathbf{V}$ are randomly initialized while satisfying the transmit power constraint.\footnote{Although the initial digital beamformers for the information signal and AN are randomly generated, the proposed algorithm consistently converges to a stable solution.}
VRs for all NF paths are assumed to have equal size. Following a typical non-overlapping VR assumption in \cite{ali2019linear}, the baseline configuration uses a VR size of 32 antennas, ensuring non-overlapping VRs for the $K=3$ ER LoS paths and one NF scatterer path, collectively spanning the array.

Fig.~\ref{fig2} illustrates the sum secrecy rate (SSR) versus the energy harvesting requirement $Q_0$ for $K=3$ and $K=5$ ERs, comparing the proposed scheme (for Type-I and Type-II IRs) against two benchmarks: 1) \textbf{Fully-digital scheme:} it assumes $N_{\text{RF}}=N$, replacing $\mathbf{F}_{\text{A}}\mathbf{w}_m$ and $\mathbf{F}_{\text{A}}\mathbf{v}_g$ with fully-digital beamformers $\mathbf{w}_{\text{FD,}m}\in \mathbb{C}^{N\times 1}$ and $\mathbf{v}_{\text{FD,}g}\in \mathbb{C}^{N\times 1}$, and it serves as a performance upper bound; 2) \textbf{No-AN scheme:} it sets $\mathbf{V} = \mathbf{0}$ within the proposed HB architecture. First, the SSR of all schemes decreases as $Q_0$ increases. This is because a stricter energy constraint forces the BS to allocate more power towards the ERs via AN or by adjusting information beams, which reduces the flexibility to maximize the secrecy rate for IRs. Second, Type-II IRs achieve higher SSR due to AN cancellation, but the performance gap from Type-I is relatively small. This is attributed to the NF beam focusing effect: AN directed towards NF ERs disperses significantly upon reaching FF IRs, limiting its impact. Third, the proposed scheme significantly outperforms No-AN, demonstrating AN's crucial role. Notably, at $Q_0 = 0.03\ \upmu \text{W}$ ($K=3$), the proposed scheme yields $\approx$94\% of the fully-digital SSR using only $\approx$8\% ($N_{\text{RF}}/N$) RF chains, highlighting a favorable trade-off between the performance and complexity. A key takeaway from Fig.~\ref{fig2} is the limited benefit of Type-II receivers in this setup, primarily due to the spatial separation and distinct propagation characteristics (NF vs. FF) of ERs and IRs, combined with NF beam focusing of AN.

Fig.~\ref{fig3} investigates the impact of VR size on SSR, comparing with \textbf{full-VR} (all antennas visible). When VRs are non-overlapping ($<32$ antennas), increasing size improves SSR. Larger VRs imply better effective channels, allowing the energy constraint $Q_0$ constraint to be met more efficiently and freeing power for information transmission. When VRs overlap ($>32$ antennas), SSR improves more rapidly. This is because, in addition to channel improvement, ERs receive stronger AN signals intended for adjacent ERs. This increased AN interference degrades eavesdropping capability, further boosting secrecy. The performance gap between the proposed scheme and the full-VR scheme diminishes as the VR size increases, converging when the VRs span the entire array. Overall, larger VRs enhance secrecy, particularly when they overlap, by improving energy harvesting efficiency and simultaneously increasing the effectiveness of AN against potential eavesdroppers.

Finally, Fig.~\ref{fig4} examines the influence of the IR's angular position on SSR. We consider $M=1$ IR at $1.5d_{\text{R}}$ and $K=3$ ERs symmetrically at $\theta _k\in \{ 1.6, 0, -1.6 \}$ rad, $r_k\in \{ 0.3, 0.1, 0.3 \}d_{\text{R}}$. The IR's angle is varied. We compare the proposed scheme (mixed-field) with a baseline \textbf{FF scheme}, where both the IR and ERs are in the FF. In the FF scheme, the IR (Type-I) is placed as in the mixed-field case, while the ERs maintain the same angles but are moved to the FF at distances $r_k\in \left\{ 1.3, 1.1, 1.3 \right\}d_{\text{R}}$. As expected, the secrecy rates of all schemes degrade significantly when the IR's angle approaches that of an ER. This is due to the increased spatial correlation between the IR's intended signal and the ER's channel, leading to higher signal leakage. 
An interesting observation arises from the comparison: the mixed-field scheme shows a sharp SSR decline when the IR angle is within $\approx \pm 0.4$ radians of the ERs at $\pm 1.6$ rad. In contrast, the FF scheme's decline occurs within a narrower range of $\approx \pm 0.2$ radians.
This demonstrates that the correlation between the IR's intended signal and the NF ER's channel remains significant at larger angular separations in the mixed-field scenario compared to the purely FF case. This finding aligns with \cite{zhang2023mixed}, highlighting distinct spatial characteristics and potential security vulnerabilities in mixed NF/FF scenarios. Accordingly, ensuring sufficient angular separation between the FF IRs and NF ERs in practical deployments is essential to mitigate information leakage and enhance secrecy performance.

\section{Conclusion}
This letter addressed PLS in mixed-field SWIPT with NF ERs and FF IRs. We proposed an AN-aided secure HB algorithm to maximize the WSSR for IRs with Type-I/II receivers. Simulation results demonstrated that the proposed scheme significantly enhances secrecy performance compared to relevant benchmarks, achieving this improvement with substantially lower hardware complexity. Furthermore, the results highlighted a key characteristic of ELAA-SWIPT systems in mixed-field: secrecy performance can degrade substantially even with considerable angular separation between IRs and ERs, which is a distinct behavior compared to purely FF systems.

\bibliographystyle{IEEEtran}
\bibliography{Zhang_WCL}

\end{document}